\documentclass[epj]{svjour}

\usepackage{epsfig}
\usepackage{epsf}
\usepackage[english]{babel}
\usepackage{latexsym}
\usepackage{amsmath}
\usepackage{amssymb}
\usepackage{graphicx}

\newcommand{\be}{\begin{equation}}
\newcommand{\ee}{\end{equation}}
\newcommand{\bea}{\begin{eqnarray}}
\newcommand{\eea}{\end{eqnarray}}
\newcommand{\h}{\mathcal{H}}
\newcommand{\ave}[1]{\langle #1 \rangle}

\begin{document}

\title{Limited resolution in complex network community detection with Potts model approach}
\author{Jussi M. Kumpula\inst{1}  \and Jari Saram\"{a}ki\inst{1} \and Kimmo Kaski\inst{1}  \and J\'{a}nos Kert\'{e}sz\inst{1,2}}
\institute{Laboratory of Computational Engineering, Helsinki University of Technology, P.O. Box 9203, FIN-02015 HUT, Finland \\ 
\mail{jkumpula@lce.hut.fi}
  \and  Department of Theoretical Physics, Budapest University of Technology and Economics, Budapest, Hungary}

\date{\today}

\abstract{
According to Fortunato and Barth\'elemy, modularity-based community detection
algorithms have a resolution threshold such that small communities in a large
network are invisible. Here we generalize their work and show that the $q$-state
Potts community detection method introduced by Reichardt and Bornholdt 
also has a resolution threshold. The model contains a parameter by which this threshold
can be tuned, but no {\it a priori} principle is known to select the proper value.
Single global optimization criteria do not seem capable for detecting all
communities if their size distribution is broad.
\PACS{
  {89.75.-k}{Complex systems} \and
  {89.75.Hc}{Networks and genealogical trees} \and
  {89.75.Fb}{Structures and organization in complex systems} \and
  {89.65.-s}{Social and economic systems}
}
}

\maketitle

\selectlanguage{english}

\section{Introduction}

Networks are an efficient way to represent a variety of complex systems,
including technological, biological and social systems \cite{RefWorks:72,RefWorks:77}. 
Many networks have substructures called communities, which
are, loosely speaking, groups of nodes that are densely interconnected but only
sparsely connected with the rest of the network  \cite{RefWorks:76,RefWorks:75,RefWorks:67,RefWorks:52}. 
Detecting such communities is of interest, because they may provide 
valuable information of the substructure and functionality of the network, e.g., 
functional modules in metabolic networks,
communities of individuals interacting with each other, {\it etc}.
This analysis can also be extended to more complex properties,
including networks of communities \cite{RefWorks:81}, roles of 
nodes inside and between communities \cite{RefWorks:52}, 
and the effect of communities on the dynamics of for example information flow 
through the network \cite{RefWorks:82}. 

A large number of algorithms have been developed for detecting the communities, 
for reviews see \cite{RefWorks:53,RefWorks:70}. A particularly popular method is based 
on the concept of modularity $Q$ introduced by Newman and Girvan \cite{RefWorks:46}:
\be
Q =  \sum_{s} e_{ss}-a_s^2,
\label{eq:modularity}
\ee
where $e_{rs}$ is the fraction of links that fall between nodes
in communities $r$ and $s$ and $a_s = \sum_r e_{rs}$. Detecting communities is 
equivalent to optimizing the modularity of the network, where optimization is 
computationally demanding, especially for large networks, but solvable with various 
approximate methods \cite{RefWorks:40,RefWorks:49,RefWorks:46,RefWorks:68,RefWorks:65}. 
Modularity optimization has been shown to perform well for many 
test networks \cite{RefWorks:53,RefWorks:69}. 

Recently, Fortunato and Barth\'{e}lemy showed that modularity optimization fails to find
small communities in large networks, indicating that it is favorable to 
combine small communities into larger ones  \cite{RefWorks:55}. 
In a network which has $L$ links, there is a characteristic number of links, such that
communities with less than $\sqrt{L/2}$ links are not visible. Earlier Reichardt and Bornholdt 
(RB) had introduced a general framework for community detection, which includes the 
modularity optimization as a special case  \cite{RefWorks:34,RefWorks:50}.
Starting from a $q-$state Potts Hamiltonian, they show that community detection 
can be interpreted as finding the ground state of an infinite-range spin-glass. Potts spins 
are assigned to the nodes of the network and  the communities can be identified as clusters 
of aligned spins in the ground state. The model is based on a comparison of the investigated 
network to a null model which can be arbitrarily chosen. In addition, the method contains 
a tunable parameter $\gamma$ for detecting community structures at different hierarchical levels. 
The Newman-Girvan modularity optimization method is a special case in this general 
framework, where the null model is the configuration model \cite{RefWorks:73} 
and $\gamma =1$. The question arrises whether the more general RB spin-glass-based community 
detection method is able to overcome the limitations of the modularity optimization. 
Our paper addresses this question.

We analyze the
effect of $\gamma$ on community detection and consider how to design a network with 
optimal community structure, study the resolution limit and its estimates by using a general 
null model, and, finally, demonstrate the consequences of our findings in certain example cases.

\section{Optimal number of communities in the RB model}
\label{sec:gamma}

For detecting the communities in a network, Reichardt and Bornholdt
proposed the following Hamiltonian:
\be
\h = - \sum_{i\neq j} \left( A_{ij} - \gamma p_{ij} \right) \delta(\sigma_i,\sigma_j),
\label{eq:pottsH1}
\ee
where $A_{ij}$ denotes the adjacency matrix of the graph with $A_{ij}=1$
if an edge is present and zero otherwise, $\sigma_i \in \{ 1,2,\ldots,q \}$ denotes
the group index of node $i$, $\gamma$ is a parameter of the model, and 
$p_{ij}$ denotes the link probability between nodes $i$ and $j$ according to the null 
model. The null model reflects the connection probability between nodes in a network 
having no apparent community structure. Possible choices for the null model are, 
for example, $p_{ij}=p$ and $p_{ij}=\frac 1 {2L} k_i k_j$, 
where $k_i$ is the degree of node $i$ and $L$ is the number of links in the network. 
The former null model corresponds to the Erd\"os-R\'enyi network \cite{RefWorks:71}, 
whereas the latter one is closely related to the configuration model.
The Hamiltonian (\ref{eq:pottsH1}) rewards existing links inside communities, but the 
reward is reduced if $p_{ij}$ is large. Furthermore,
the penalty of a missing link inside a community is proportional to its probability. 
The modularity $Q$ of Eq.~(\ref{eq:modularity}) is related to Eq.~(\ref{eq:pottsH1}) as
$Q=-\h/L$, provided that $\gamma=1$ and $p_{ij}=\frac 1 {2L} k_i k_j$.

In order to gain some insight to the model given by Eq.~(\ref{eq:pottsH1}),
we consider two limits of $\gamma$. First, when $\gamma \to 0$ each link inside a community 
comes as a ``surprise'', while the missing links are not increasing the energy as they are
not expected to exist. Thus, in the limit $\gamma=0$ the minimum energy is obtained
when all nodes are assigned into the same community, and the minimum energy is $\h=-2L$. 
Second, when $\gamma \gg 1$ communities are broken into smaller pieces because
the penalty from missing links is large and all existing links are considered to be extremely 
likely. When $\gamma$ exceeds the inverse of the minimum of non-zero $p_{ij}$:s, the 
terms $A_{ij}-\gamma p_{ij}$  in (\ref{eq:pottsH1}) become all negative, and the minimum 
energy is obtained when each node is regarded as a separate community, resulting in $\h = 0$.
This demonstrates that for small values of $\gamma$, one can expect to find large community 
structures, whereas for large values of $\gamma$ only small community structures are found.
The total amount of energy that can possibly be contributed by links and non-links is equal 
for $\gamma=1$, which can be regarded as a natural choice. Later we show, however, that 
optimizing the energy with $\gamma=1$ does not necessarily yield the obvious and most 
natural community structure even in a simple test case.

\begin{figure}[t]
\includegraphics[width=0.9\linewidth]{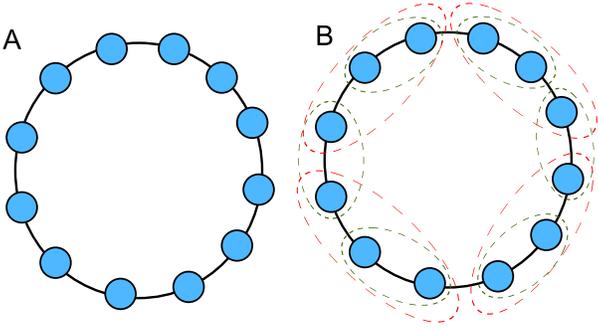}
\caption{A: a ring-like network of $n$ cliques joined by $n$ links. 
B: Consecutive cliques can be merged to form larger communities. The optimal configuration
depends on the network parameters and  $\gamma$. }
\label{fiq:cliqueRing}
\end{figure}

Following the steps in \cite{RefWorks:55}, we next consider how to design
a connected network with $N$ nodes and $L$ links such that the energy (\ref{eq:pottsH1})
is minimized. 
In particular, we are interested in the optimal number
of communities as a function of $L$ and $\gamma$.
Therefore, we study a network which has $\hat{n}$ fully connected subgraphs (or cliques) 
of equal size, being interconnected with
$\hat{n}$ links and arranged in a ring-like structure, see Fig.~\ref{fiq:cliqueRing}(A). 
This network has by construction $\hat{n}$ communities, namely the cliques, i.e., the links 
inside the cliques are intra-community, while those connecting them are inter-community links. 
The minimization of (\ref{eq:pottsH1}) should reflect this structure providing the $\hat{n}$ 
equal size communities. Moreover, for such an obvious structure this result should be robust 
against changing $\gamma$ or even the null model.

Equation (\ref{eq:pottsH1}) can be rewritten as
\be
\h = - \sum_{s=1}^n \left( l^s -\gamma [l]^s_{p_{ij}} \right),
\label{eq:pottsH2}
\ee
where $l^s$ is the number of links inside community $s$ and $[l]^s_{p_{ij}}$ is the 
expected number of links in that community given the link distribution
$p_{ij}$ and the current assignment of nodes into communities \cite{RefWorks:50}.
In order to be compatible with the calculations in \cite{RefWorks:55}, 
we choose first to use $p_{ij}=\frac{1}{2L} k_i k_j$, i.e., our reference system is the configuration model. 
In this case, $[l]^s_{p_{ij}}=\frac{1}{4L}K_s^2$, where $K_s$ is the sum
of degrees of nodes in community $s$. 
It is straightforward to show that Eq.~(\ref{eq:pottsH2})
is minimized when each community has $L/n-1$ links. Then, the energy is
\be
\h_{min}(n,\gamma,L) = -(L-n-\gamma \frac L n ).
\label{eq:minE1}
\ee
The optimal number of communities, $n^*$, is obtained as the zero
of the derivative $d\h_{min}(n,\gamma,L)/dn$. This yields $n^*=\sqrt{\gamma L}$, 
which in turn gives back the result of \cite{RefWorks:55} for $\gamma =1$.
If the null model is  $p_{ij}=p$, i.e., an Erd\H os-R\'enyi graph, a similar 
calculation shows that the energy minimum is obtained when each community has 
an equal number of nodes. In this case, the optimal number of communities 
is $n^*=\sqrt{\gamma L \frac N {N-1}}$.

Let us suppose that, given $N$ and $L$, we have constructed a ring-like network 
as described above, having more than $\sqrt{\gamma L}$ cliques. 
Previous analysis shows, counterintuitively, that when each clique is
considered as a separate community the energy (\ref{eq:pottsH2}) is not minimized. 
Instead, it is better to relabel the communities so that small communities are 
merged to form larger ones. On the other hand, if the number of 
communities is much smaller than $\sqrt{\gamma L}$ it might be advantageous
to split large communities into smaller ones.
Therefore, the original, well defined communities are not necessarily
found by optimizing the quality function (\ref{eq:pottsH2}).
In particular, small communities may remain unresolved. 

\section{Resolution threshold with a general null model}
\label{sec:genNullMod}

The previous section suggests that the most
common null models, $p_{ij}=\frac{1}{2L} k_i k_j$ and $p_{ij}=p$, 
lead to merging of small communities in large networks. 
In this section we investigate the case of a general null model and the effect
of $\gamma$ on the resolution. 
Hence, we consider a general undirected, unweighted network having $N$ nodes and $L$ links. 
Let us suppose that the nodes have somehow been assigned to communities.
We take two communities, labeled $s$ and $r$,
each having $l^s$ and $l^r$ links inside and $l^{s\leftrightarrow r}$ links between them.
The question is, when should the communities be merged?
At first, the energy (\ref{eq:pottsH2}) reads as follows 
\be
E_1 = \sum_{t\neq s,r} ( -l^t + \gamma [l]_{p_{ij}}^t ) + (-l^r+\gamma [l]_{p_{ij}}^r) +  (-l^s+\gamma [l]_{p_{ij}}^s)
\ee
whereas after combining the communities the energy is
\be
E_2 = \sum_{t\neq s,r} ( -l^t + \gamma [l]_{p_{ij}}^t ) + \left[ -(l^r+l^s+l^{s \leftrightarrow r}) + \gamma [l]_{p_{ij}}^{s+r} \right],
\ee
where $[l]_{p_{ij}}^{s+r}$ is the expected number of links in the combined community and
$l^{s\leftrightarrow r}$ is the number of links between the communities $s$ and $r$.
The communities should be combined if
\be
\Delta E = E_2-E_1 = -l^{s \leftrightarrow r} + \gamma \left( [l]_{p_{ij}}^{s+r} - [l]_{p_{ij}}^r -[l]_{p_{ij}}^s \right) < 0.
\label{eq:deltaE}
\ee
But $[l]_{p_{ij}}^{s+r} - [l]_{p_{ij}}^r -[l]_{p_{ij}}^s \equiv [l]_{p_{ij}}^{s\leftrightarrow r}$ is the expected number of links between the 
communities 
and equation (\ref{eq:deltaE}) reduces to
\be
\gamma [l]_{p_{ij}}^{s\leftrightarrow r} < l^{s \leftrightarrow r}.
\label{eq:gammaCriterion}
\ee
As the communities have  $n_s$ and $n_r$ nodes each,
the maximum number of links between the communities is $n_s n_r$. 
In a large network, 
the average probability for connecting two nodes has to 
be of the order of $N^{-1}$, regardless of the null model. 
Therefore, the expected number of 
links between the communities, $[l]_{p_{ij}}^{s\leftrightarrow r}$, is on average of the 
order of $ n_s n_r / N$.
Using this estimate in Eq.~(\ref{eq:gammaCriterion}) suggests that even a single link 
between small communities may trigger merging if the communities are small, i.e., $n_s, n_r \ll N$. 
In particular, communities of approximately the same size are merged if
\be
n_s \approx n_r \lesssim \sqrt{Nl^{s\leftrightarrow r}/\gamma}.
\label{eq:resLim1}
\ee
Now, let us suppose the communities are loosely connected to each other, that is, $l^{s\leftrightarrow r} \sim 1$.
When this is applied in Eq.~(\ref{eq:resLim1}), we obtain that it is
beneficial to combine communities smaller than $\sim \sqrt{N/\gamma}$. 
This is the lower limit for the community size that the method is able to detect.
Large values of $\gamma$ decrease this resolution threshold, but rather inefficiently.
When the communities are more densely interconnected, the resolution threshold increases.
In the extreme (unphysical) limit, when the communities are connected with $l^{s\leftrightarrow r} \sim L$ links, Eq.~(\ref{eq:resLim1}) indicates that even communities whose size is comparable to the whole network
may remain unresolved.
Similar results for the resolution thresholds were obtained 
in Ref.~\cite{RefWorks:55} for the case  $\gamma=1$, $p_{ij}=k_i k_j/2L$:
Two tightly connected communities may be merged if each has less than $L/4$ links, whereas
the lower limit is $\sqrt{L/2}$ for communities connected with a single link.

The community structure found by the RB model corresponds to the global minimum of (\ref{eq:pottsH2}).
It should be noted that the previous calculations do not prove that the {\it particular} communities
$s$ and $r$ will be in the same community for the global minimum.
The calculations show, however, that the global minimum does not contain 
connected communities smaller than the above mentioned size limits,
because by combining them a lower energy would be achieved.

Equation (\ref{eq:gammaCriterion}) shows also that cliques are stable against splitting for any reasonable $\gamma$. Suppose that a clique is split into two parts each having $n_s$ and $n_r$ nodes. 
The parts have the maximum number $l^{s\leftrightarrow r}=n_s n_r$ of connecting links.
Substituting this and $[l]_{p_{ij}}^{s\leftrightarrow r} \sim n_s n_r / N$ into Eq.~(\ref{eq:gammaCriterion})
shows that it is beneficial to split a clique only when $\gamma \sim N$. 
Such high value of $\gamma$ does not, however, make sense because it would 
lead to splitting the network into individual nodes for the following reason.
In this case the average value of links from a node according to the null model would exceed the maximum possible number of links from 
a node, and the communities would be split into individual nodes.
We conclude that when $\gamma \ll N$ cliques and almost complete cliques are not split.

\section{Examples}
\label{sec:examples}

\begin{figure}[t]
\includegraphics[width=0.8\linewidth]{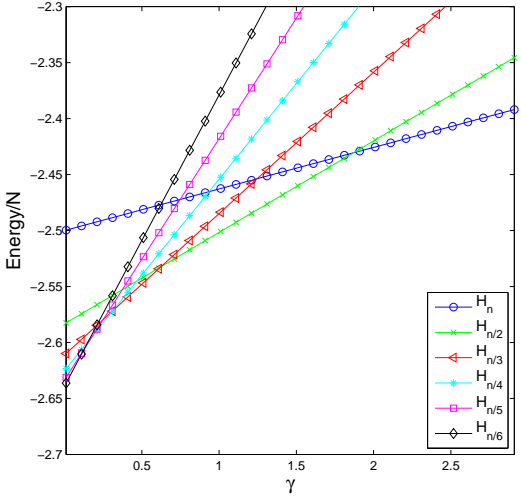}
\caption{Energy (\ref{eq:cliqueEnergy2}) as a function of $\gamma$ for a system where $n=60$, $m=6$ and $r=1,\ldots,6$.
The optimal configuration depends on $\gamma$ and the natural communities are found only when
$\gamma > 1.875$, c.f. Eq.~(\ref{eq:gammaLimit}).}
\label{fiq:minEs}
\end{figure}

We illustrate the consequences of the above results in three example cases.
Let us first consider the simplest possible
case of community detection \cite{RefWorks:55}:  
the network consists of a ring of complete cliques joined by single links, 
Fig.~\ref{fiq:cliqueRing}(A).
There are $n$ cliques and each clique has $m$ nodes and $m(m-1)/2$ links.
Figure \ref{fiq:cliqueRing}(B) shows a case where
$r$ consecutive cliques are merged to form a single community. 
A straighforward calculation shows that in this case, the energy is given by
\be
\h_{\frac n r}  (\gamma)  = -n \left(\frac {m(m-1)} 2 + \frac {r-1} r\right) + \gamma \frac {rn} {4L} m^2 (m-1)^2,
\label{eq:cliqueEnergy2}
\ee
when $p_{ij}=\frac{1}{2L} k_i k_j$.
By joining cliques, we get a ``bonus'' from the links joining the cliques, i.e., term $(r-1)/r$, 
but in large communities the expected number of links inside the communities
is increasing faster than in small communities. Thus, 
for small $\gamma$ the merged cliques have low energy, 
but as $\gamma$ increases the energy is growing quite fast as illustrated in Fig.~\ref{fiq:minEs}.
The optimal configuration found by optimizing Eq.~(\ref{eq:pottsH1}) is 
the configuration that has the lowest energy
for the given values of $n$, $m$ and $\gamma$. 
Especially, it can be shown that the natural communities are found only if 
\be
m(m-1) + 2 > \frac n \gamma.
\label{eq:gammaLimit}
\ee
When the link probability is $p_{ij}=p$, we obtain the same result with a 
correction term of the order of $(\gamma m)^{-1}$.

Our second example is a random network, which has often been used as a test network for 
community detection algorithms \cite{RefWorks:46}. 
The network consists of $n$ communities each having $m$ nodes.
Each node has on average $\ave k$ links of which $\ave{k_{in}}$ go to random nodes in 
the same group and
$\ave{k_{out}}=\ave k - \ave{k_{in}}$ links 
lead randomly to nodes in other communities.
Let us now calculate when, on the average, it is beneficial to merge two designed communities.
We obtain that the average number of observed links between the communities is
\be
\ave {l^{s\leftrightarrow r}}=\frac m {n-1} \ave{k_{out}},
\label{eq:linkNum}
\ee 
where the averaging is done over all the realizations of the network. 
Note that if 
 $m/(n-1) \ave{k_{out}} < 1$ we have to set $\ave {l^{s\leftrightarrow r}}=1$ because 
we are considering only communities which are connected by at least one link.
The null model is again $p_{ij}=\frac{1}{2L} k_i k_j$.
According to the null model the expected number of links between communities
is 
\be
[l]_{p_{ij}}^{s \leftrightarrow r}=\frac 1 {2L} (m\ave k)^2=\frac {m\ave k} n,
\label{eq:ex2_2}
\ee
when averaged over the realizations of the network.
Now Eqs.~(\ref{eq:gammaCriterion}), (\ref{eq:linkNum}) and (\ref{eq:ex2_2}) 
give that the communities are merged if
\be
\gamma < \left\{ \begin{array}{ll}
 \frac {\ave{k_{out}}} {\ave k} \frac {n}{n-1} & \textrm{for } n < 1 + m\ave{k_{out}} \\
\frac {n}{m \ave k} &   \textrm{for } n > 1 + m\ave{k_{out}}.
\end{array} \right.
\label{eq:newmanPairs2}
\ee
For typical values $n=4$, $m=32$, $\ave k=16$ and $\ave{k_{out}}=1\ldots8$ we find that $\gamma=1$
should give the correct communities. 
Thus, it is not surprising that community detection based  
modularity optimization (\ref{eq:modularity})  performs well for this network. 
We point out that it is possible to choose the parameters $n$, $m$, $\ave k$ and $\ave {k_{out}}$ 
in such a way that modularity optimization with $\gamma=1$ does not give the designed communities.

As a third example we note that
the Potts Hamiltonian (\ref{eq:pottsH1}) can be generalized to weighted networks by using
a weighted adjacency matrix $W_{ij}$. A simple way to do this is to define
\be
\h_w = - \sum_{i\neq j} (W_{ij}-\gamma \overline{w_{ij}} p_{ij})\delta (\sigma_i,\sigma_j),
\ee 
where $ \overline{w_{ij}} $ is the average link weight. In this way, strong links inside
communities lower the energy greatly, while missing links are assumed to be of average weight.
Using weights does not, however, resolve the underlying problem that in a large
network even a single link easily exceeds the expected weight between the communities.

\begin{figure}[t]
\includegraphics[width=1.0\linewidth]{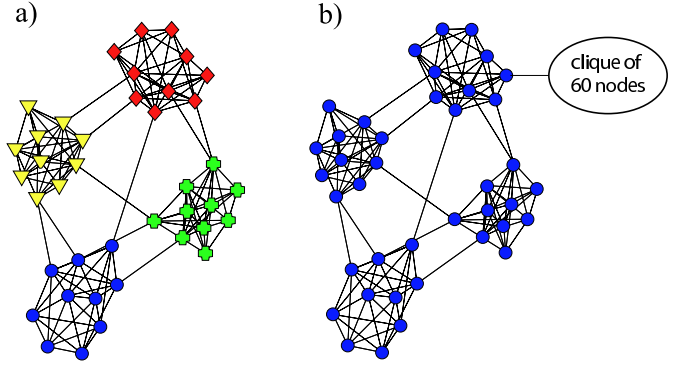}
\caption{An example of the effect of network size on the resolution of the Potts method. Symbols correspond to communities. See text for details.}
\label{4clique_fig}
\end{figure}

Finally, in Figure \ref{4clique_fig} we demonstrate the effect of network size on the resolution 
of the Potts method. Panel a) shows a network of four groups of 10 nodes. We have compared the 
energies (3) for two community divisions using $\gamma=1$ and the configuration null model. 
$E_1=0$ is the energy for the case when all four groups are assigned to a single community, 
whereas $E_4=-100.2$ is the energy when the four groups are each assigned to a different community. 
In this case, $E_4<E_1$, i.e. the groups are properly identified as communities. However,
if the original network of panel a) is modified such that an additional 60-clique community 
is connected to it via a single link, the situation is changed. All nodes of this new 60-clique 
are assigned to a single community. Now, $E'_1=-271.17$ is the energy when the original four groups
are merged into a Potts community, and $E'_4=-269.73$ the energy when
they are assigned to separate communities. Hence $E'_1<E'_4$, i.e., the energy for merged groups 
is lower. This is unphysical, since connecting the new clique via a single link does not alter the 
original four-group topology.

\section{Conclusions}

In the light of the above considerations it is clear that the problem of the resolution 
limit is not restricted to the Newman-Girvan method of modularity optimization. Rather, it is
a flaw which seems to be present in any community detection scheme based on global 
optimization of intra- and extra-community links and on a comparison to any null model. 
The limited resolution rises from the fact that in a large network the expected number of 
links between two small sets of nodes is small and even a single link between the sets is 
enough to merge them. The null model uses the global probability of connecting nodes 
while the small communities should be considered on a more local level. We agree with the 
conclusion of Ref.~\cite{RefWorks:55} that presently, in large networks, local community 
detection methods like \cite{RefWorks:52} seem to perform better from
the point of view of resolution. An alternative solution to this 
problem could be to iteratively change the parameter $\gamma$ when 
looking for smaller communities in a large network.

Our results indicate that when the community structure is not known beforehand, there is 
no simple way to decide which $\gamma$ gives the most relevant communities. Moreover, if the 
size distribution of the communities is broad, like in collaboration networks 
\cite{RefWorks:52} or school friendship networks \cite{GHKV}, there is no single proper 
value of $\gamma$ for the optimal resolution. The hierarchical structure can be examined 
to some extent by using several values of $\gamma$ \cite{RefWorks:50}, but this method may find too 
much hierarchy in the network as it tends to artificially merge communities. 
Because of this tendency, one should always carefully 
investigate the structure of the found communities. 

{\bf Acknowledgements:} JK thanks Santo Fortunato for inspiring discussions at ISI, Torino. 
This work was partially supported by OTKA K60456 and the Academy of Finland (Center of Excellence program 2006-2011).


\begin{thebibliography}{21}

\bibitem{RefWorks:72}
R.~Albert, A.L. Barab\'{a}si, Rev. Mod. Phys. \textbf{74}, 47 (2002)

\bibitem{RefWorks:77}
S.~Boccaletti, V.~Latora, Y.~Moreno, M.~Chavez, D.U. Hwang, Phys. Rep.
  \textbf{424}, 175 (2006)

\bibitem{RefWorks:76}
R.~Guimer\'a, S.~Mossa, A.~Turtschi, L.A.N. Amaral, PNAS \textbf{102}, 7794
  (2005)

\bibitem{RefWorks:75}
A.~Arenas, L.~Danon, A.~D\'{i}az-Guilera, P.M. Gleiser, R.~Guimer\'{a}, Eur.
  Phys. J. B \textbf{38}, 373 (2004)

\bibitem{RefWorks:67}
R.~Guimera, L.A.N. Amaral, Nature \textbf{433}, 895 (2005)

\bibitem{RefWorks:52}
G.~Palla, I.~Der\'enyi, I.~Farkas, T.~Vicsek, Nature \textbf{435}, 814 (2005)

\bibitem{RefWorks:81}
P.~Pollner, G.~Palla, T.~Vicsek, Europhys. Lett. \textbf{73}, 478 (2006)

\bibitem{RefWorks:82}
J.P. Onnela, J.~Saram\"aki, J.~Hyv\"onen, G.~Szabo, D.~Lazer, K.~Kaski,
  J.~Kert\'{e}sz, A.L. Barab\'{a}si, Proc. Natl. Acad. Sci. (USA), in press, e-print physics/0610104

\bibitem{RefWorks:53}
L.~Danon, A.~D\'{i}az-Guilera, J.~Duch, A.~Arenas, J. Stat. Mech.
  \textbf{2005}, P09008 (2005)

\bibitem{RefWorks:70}
M.E.J. Newman, Eur. Phys. J. B \textbf{38}, 321 (2004)

\bibitem{RefWorks:46}
M.E.J. Newman, M.~Girvan, Phys. Rev. E. \textbf{69}, 026113 (2004)

\bibitem{RefWorks:40}
M.E.J. Newman, Phys. Rev. E \textbf{74}, 036104 (~19) (2006)

\bibitem{RefWorks:49}
A.~Clauset, M.E.J. Newman, C.~Moore, Phys. Rev. E \textbf{70}, 66111 (2004)

\bibitem{RefWorks:68}
J.~Duch, A.~Arenas, Phys. Rev. E \textbf{72}, 027104 (2005)

\bibitem{RefWorks:65}
M.E.J. Newman, Phys. Rev. E \textbf{69}, 066133 (2004)

\bibitem{RefWorks:69}
M.~Gustafsson, M.~Hornquist, A.~Lombardi, Physica A \textbf{367}, 559 (2006)

\bibitem{RefWorks:55}
S. Fortunato, M.~Barth\'elemy, Proc. Natl. Acad. Sci. USA {\bf 104}, 36-41 (2007), e-print physics/0607100

\bibitem{RefWorks:34}
J.~Reichardt, S.~Bornholdt, Phys. Rev. Lett. \textbf{93}, 218701 (2004)

\bibitem{RefWorks:50}
J.~Reichardt, S.~Bornholdt, Phys. Rev. E \textbf{74}, 016110 (2006)

\bibitem{RefWorks:73}
M.E.J. Newman, SIAM Review \textbf{45}, 167 (2003)

\bibitem{RefWorks:71}
P.~Erd\"os, A.~R\'enyi, Publ.Math.Debrecen \textbf{6}, 290 (1959)

\bibitem{GHKV}
M.C. Gonzales, H.J. Herrmann, J. Kert\'esz, T. Vicsek, Physica A (in press), physics/0611268

\end{thebibliography}

\end{document}